\documentclass[fleqn,10pt]{wlscirep}

\usepackage{wrapfig}
\usepackage{subfigure}
\usepackage[outdir=./]{epstopdf}
\usepackage[export]{adjustbox}

\usepackage{gensymb}
\usepackage{CJKutf8}

\renewcommand{\vec}[1]{\mathbf{#1}} 
\newcommand{\mat}[1]{\mathbf{#1}} 

\title{Toward Creating Subsurface Camera}

\author[1,*]{WenZhan Song}
\author[1]{Fangyu Li}
\author[1]{Maria Valero}
\author[2]{Liang Zhao}
\affil[1]{Center for Cyber-Physical Systems, University of Georgia}
\affil[2]{Division of Math and Computer Science, University of South Carolina Upstate}

\affil[*]{Email: wsong@uga.edu}

\begin{abstract}
In this article, the framework and architecture of Subsurface Camera (SAMERA) is envisioned and described for the first time. A SAMERA is a geophysical sensor network that senses and processes geophysical sensor signals, and computes a 3D subsurface image in-situ in real-time. The basic mechanism is: geophysical waves propagating/reflected/refracted through subsurface enter a network of geophysical sensors, where a 2D or 3D image is computed and recorded; a control software may be connected to this network to allow view of the 2D/3D image and adjustment of settings such as resolution, filter, regularization and other algorithm parameters. System prototypes based on seismic imaging have been designed. 
SAMERA technology is envisioned as a game changer to transform many subsurface survey and monitoring applications, including oil/gas exploration and production, subsurface infrastructures and homeland security, wastewater and CO2 sequestration, earthquake and volcano hazard monitoring. The system prototypes for seismic imaging have been built. Creating SAMERA requires an interdisciplinary collaboration and transformation of sensor networks, signal processing, distributed computing, and geophysical imaging.

\end{abstract}

\begin{document}

\flushbottom
\maketitle
%
%

\section{Introduction} 

\begin{figure}[htp]
\centering
    \includegraphics[width=0.98\textwidth]{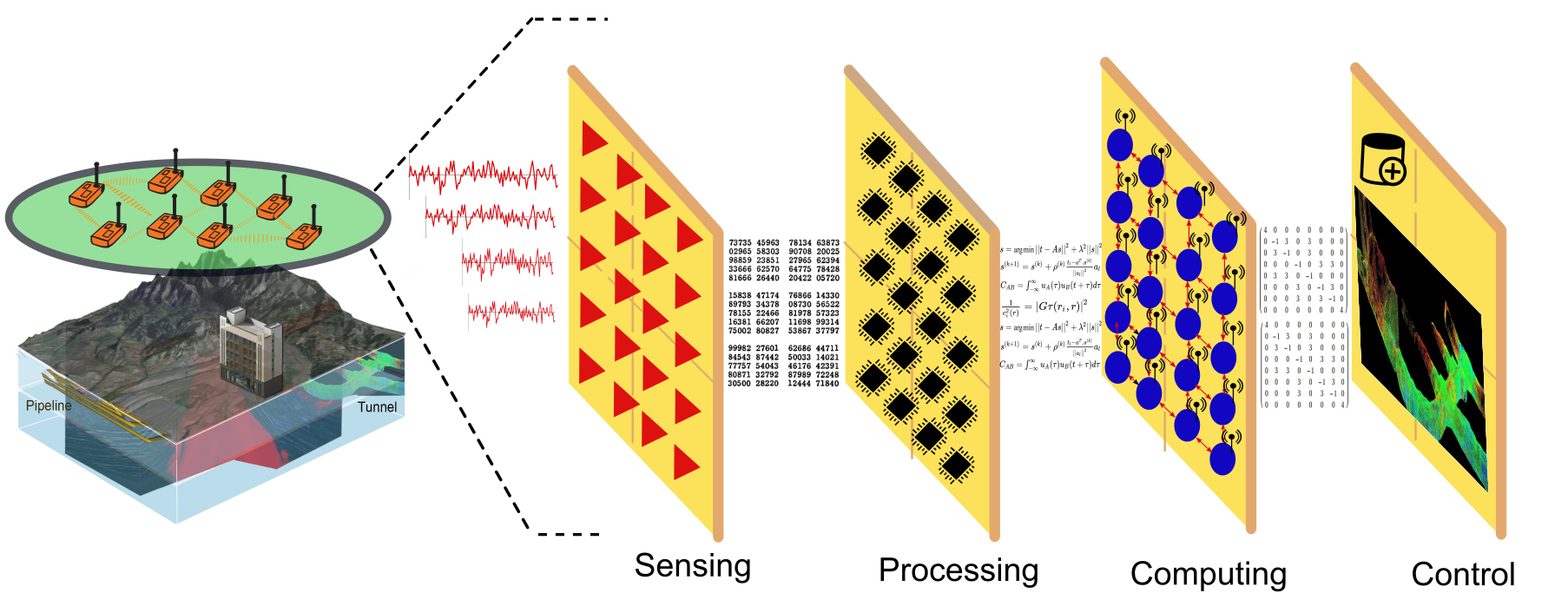}
    \caption{\textbf{SAMERA system architecture: sensing, processing, computing and control 
    }}
    \label{fig:scamera}
\end{figure}

In the eighteenth century, the concept of optical camera was conceived. The basic mechanism is: light rays reflected from a scene enter an enclosed box through a converging lens and an image is recorded on a light-sensitive medium (film or sensor); a display, often a liquid crystal display (LCD), permits the user to view the scene to be recorded and adjust settings such as ISO speed, exposure, and shutter speed. In this article, the concept of Subsurface Camera (SAMERA) is envisioned and described for the first time. The basic mechanism (Figure \ref{fig:scamera}) is as follows: geophysical waves propagating/reflected/refracted through subsurface enter a network of geophysical sensors where a 2D or 3D image is computed and recorded; a control software with graphical user interface (GUI) can be connected to this network to visualize computed images and adjust settings such as resolution, filter, regularization and other algorithm parameters. 

A SAMERA is a geophysical sensor network that senses and processes geophysical waveform signals, and computes a 3D subsurface image in-situ in real-time. Just as a camera can become a video camera to record a sequence of images, SAMERA can generate a time slice of subsurface images and enable searching, identifying and tracking underground dynamics for security and control applications. Also, just as flash lights may be added to an optical camera to enlighten the scene, geophysical transmitters may be added to a SAMERA to illuminate subsurface for faster image generation and finer resolutions, not merely relying on passive natural events (such as earthquakes). The geophysical transmitters may be add-ons to receivers (e.g., sensors) and convert receivers to transceivers. For example, seismic exploration of oil/gas industry often use explosives or vibroseis to generate active seismic waves, and GPR (Ground Penetrating Radar) equips active electromagnetic wave transmitters. Geophysical transmitters can transmit waves at different wavelengths to enable subsurface imaging at different ranges and resolutions. Waves with longer wavelength typically propagate deeper and further, but generate lower resolution images. If each geophysical transceiver is installed on mobile robots, it would enable a mobile and zoom-able SAMERA. Given an area, the geophysical transceivers can first spread out to form a sparse array to transmit waves with longer wavelength and generate a coarser subsurface image; if an interested region is identified from the coarser image, the geophysical transceivers can gather closer to form a dense array to transmit waves with shorter wavelength and generate finer subsurface images.

SAMERA technology is envisioned as a game changer to transform many subsurface survey and monitoring applications, including oil/gas exploration and production, subsurface infrastructures and homeland security, wastewater and CO2 sequestration, earthquake and volcano hazard monitoring. The system prototypes for seismic imaging have been built (section \ref{sec:prototype}). Creating SAMERA requires an interdisciplinary collaboration and transformation of sensor networks, signal processing, distributed computing, and geophysical imaging.


\section{System Framework and Architecture}
\label{sec:arch}
A SAMERA system is a general subsurface exploration instrumentation platform and may incorporate one or more types of geophysical sensors and imaging algorithms based on application needs. Various geophysical sensors and methods have been used for subsurface explorations: 
seismic methods (such as reflection seismology, seismic refraction, and seismic tomography); seismoelectrical methods; geodesy and gravity techniques(such as gravimetry and gravity gradiometry); magnetic techniques (including aeromagnetic surveys and magnetometers); electrical techniques (including electrical resistivity tomography, induced polarization, spontaneous potential and marine control source electromagnetic (mCSEM) or EM seabed logging; electromagnetic methods (such as magnetotellurics, ground penetrating radar and transient/time-domain electromagnetics, surface nuclear magnetic resonance (also known as magnetic resonance sounding)).
For the environmental engineering application, often the seismic, EM and electric resistivity methods are used. All of these geophysical imaging methods can be implemented as a type of SAMERA with the same system framework and architecture, as illustrated in Figure~\ref{fig:scamera}. 

For the simplicity of presentation and illustration, the following sections will describe SAMERA based on seismic imaging, while the framework, architecture, and algorithms will similarly apply to other geophysical sensors and imaging approaches. Choosing the seismic imaging as the example is also because seismic methods are widely used in subsurface explorations ranging from meters to kilometers in distance and depth.

\subsection{Sensing Layer and Hardware Platform}
\begin{wrapfigure}{r}{0.40\textwidth}
    \centering
    \includegraphics[width=0.99\linewidth]{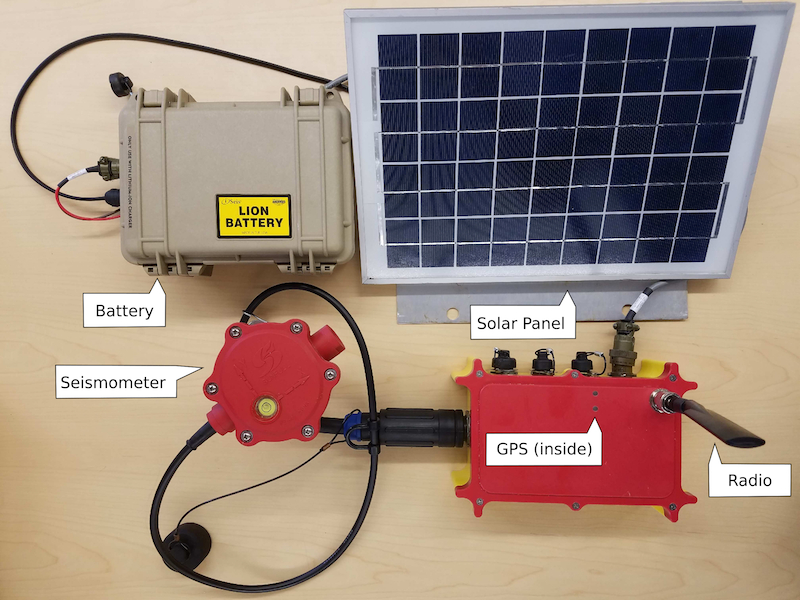}
    \caption{\textbf{SAMERA hardware platform prototype. It has geophone, GPS, computing board, wireless radio, solar panel and battery.}
    }
    \vspace{-2mm}
    \label{fig:station}
\end{wrapfigure}
\begin{figure}[htp]
\centering
    \includegraphics[width=0.98\textwidth]{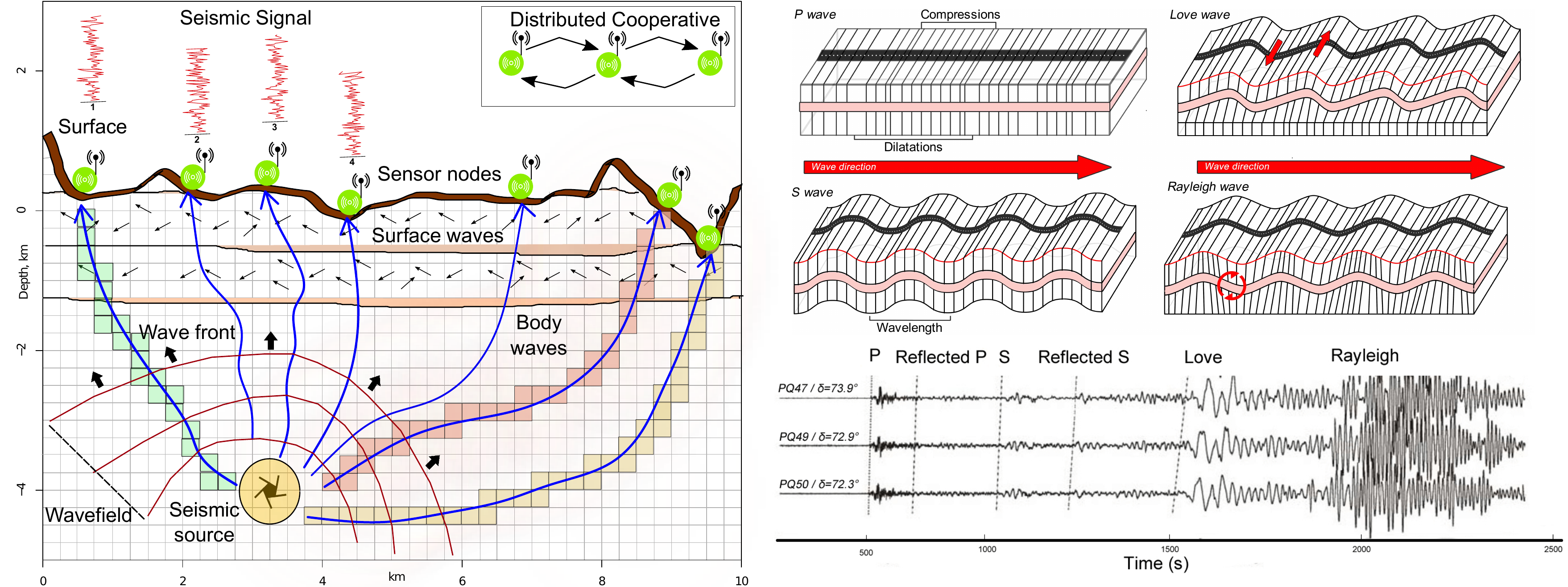}
    \caption{\textbf{Illustration of seismic imaging with body waves (P and S), surface waves (Love and Rayleigh). 
    }}
    \label{fig:waves}
\end{figure}

For seismic imaging, two types of seismic waves are mainly used in sensing layer: body waves (P and S) and surface waves (Rayleigh wave and Love wave), as illustrated in Figure \ref{fig:waves}. They have different particle movement patterns~\cite{seismicwaves}, resulting in different waveform characteristics and velocities~\cite{wilde2008passeq}. In section \ref{sec:prototype}, several seismic imaging algorithms using body and surface waves respectively will be introduced. Seismometers (often geophones) are used to sense/receive seismic waves. Some seismometers can measure motions with frequencies from 500 Hz to 0.00118 Hz. Deep and large planetary scale studies often use sub-Hz broadband seismometers, while earthquake and exploration geophysics often use 2-120 Hz geophones.
A digitizer is designed to amplify signals, suppress noises and digitize data via an Analog-to-Digital Converters (ADC) chip. The digitizer of seismic application typically has 16-32 bit resolution, with sampling rate 50-1000 Hz.  

The SAMERA hardware prototypes have been designed (Figure \ref{fig:station}) for  seismic imaging methods. It has geophone, GPS (global positioning system), computing board, wireless radio, solar panel and battery. Each sensor nodes equip wireless radio to self-form sensor networks for communication and data exchanges. Sensor networks have been successfully deployed in harsh environments \cite{song2010design,song2009air,song2008optimized,huang2012real} for geophysical surveys. The GPS provides precise timestamp and location information for each node. The computing board inside is Raspberry Pi 3 \cite{upton2016raspberry}. It has 1.2 GHz CPU, 1GB RAM and GPU for intensive local computing when needed, yet can be put in sleep for very low power consumption. Several seismic imaging methods based on this hardware platform have been built and demonstrated as described in section \ref{sec:prototype}. However, for some geophysical imaging methods (such as Full Waveform Inversion (FWI)), there are some concerns on whether they can ever be performed in sensor networks, as they requires days even months computation on IBM mainframes when this article is written. Those concerns appear to be legitimate but will gradually fade out. In 1970s, an IBM mainframe computer ran at a speed of 12.5 MHz and cost \$4.6 million. People in 1970 would similarly doubt a match box sized board (like Raspberry PI) in 2018 could be 100 times faster than a mainframe and cost only \$35. In the past three decades, the CPU frequency doubles every 18 months, as predicted by Moore's law; this trend is expected to continue in the next decade. In addition to expected hardware improvements, big data, artificial intelligence and distributed computing develop quickly as well and become more and more efficient to deal with those geophysical imaging problems.

\subsection{Processing Layer}
In the processing layer, signal processing techniques are used to process the raw time series sensor signals and extract needed information for a 2D/3D image reconstruction. The signal processing functions include data conditioning, noise cancellation, changing point detection, signal disaggregation, cross-correlation, time and frequency analysis and so on. Most signal processing tasks will be performed in each node locally, while a few processing tasks may need data exchange among nodes may, such as cross-correlation used in ambient noise imaging (section \ref{sec:ansi}).


Data conditioning is the very first step of the processing layer, which includes data interpolation, axis direction initialization, phase correction, and instrumental response removal. 
The acquired data from the field may have missing data points and the sampling rates among different types of geophysical sensors can be different, so the interpolation is necessary for the following processing steps~\cite{antoniou2016digital}. For seismology experiments, the geophones are usually placed according to the directions, N (north), E (east), and Z (vertical). Yet, for unconventional seismic explorations, the geophone axes may be randomly placed otherwise the deployment will take longer time, so an axis direction initialization may be needed based on the perforation shots from the hydraulic fracturing~\cite{maxwellbook}. In addition, the recorded seismic data may have mixed phase because of the wave propagation through the complicated media, so a phase correction is adopted to make the seismic peaks and troughs more related to the geophysical structures. Also, the instrumental response needs to be removed from seismic data~\cite{bensen2007processing}.

To enhance the signals of interest, it is common to apply a bandpass (e.g. Butterworth) filter to attenuate low and high frequency noise to improve the SNR of seismic data. 
Although the bandpass filter is incapable \cite{douglas1997bandpass} to filter out all noise components, it is still widely adopted as an initial processing step and a subsequent filter with more comprehensive performances may be included. For noise cancellation purposes, a filter is needed to reveal seismic signals embedded in random noise from the seismographs with little prior knowledge of the spectral content or temporal features ~\cite{du2000noise}. The Wiener filter~\cite{li2017automatic} is an optimal filter that minimizes the mean square error criterion, under the assumption that the signal and noise are independent. 

To extract signal event information from the filtered data is another challenge. First, signal disaggregation or wavefield separation is needed to isolate different types of seismic waves or separate different wave components~\cite{nakata2015body}. Second, to acquire the time/velocity information from all wave components, the arrivals of body waves (P and S) can be picked based on the changing point detection methods, which characterize the statical properties of the time series data~\cite{allen1978automatic}. However, for the surface waves, which are highly distorted or have strong attenuation and dispersions, the arrival picking methods fail to obtain the wanted information. Thus, cross-correlation techniques as well as temporal stacking are employed to extract the surface wave propagation properties. Based on the extracted seismic wave components, to further analyze the wave properties related to velocity, space and depth, the time-frequency analysis is usually applied to estimate different velocities associated with different frequencies to characterize different subsurface properties~\cite{Lin2009}.

\subsection{Computing Layer}
In the computing layer, the reconstruction of 2D/3D seismic images often involves computations such as linear/non-linear inversion and optimization, spatial and temporal stacking. Those computations are traditionally performed in central servers and often need data from all sensors. To implement SAMERA, a key requirement is to perform those computations in sensor networks in real-time. Thus the main research challenge is to develop distributed iterative computing algorithms under network bandwidth constraints. Song et al. \cite{song2015real,kamath2016distributed,zhao2015decentralised,kamath2017distributed,ramanan2015indigo,valero2017real,lei2016sensor,kamath2015dristi,kamath2015distributed,kamath2013component,shi2013imaging} pioneered the research on in-situ seismic imaging in distributed sensor networks. 
%
The idea is to let each node {compute in an asynchronous fashion and communicate with neighbors} only while solving 2D/3D image reconstruction problem. By eliminating synchronization point and multi-hop communication used in existing distributed algorithms, the approach can better scale to large numbers of nodes, and exhibit better resilience and stability. 
Also, randomized gossip/broadcast based iterative methods have been used. In these methods, each node asynchronously performs multiple rounds of iterations of its own data, gossips/broadcasts its intermediate result with neighbors for a weighted averaging operation. The iterative computing can be based on first-order and second-order methods (such as distributed ADMM \cite{boyd2011distributed} methods). The second-order methods expect to have a faster convergence rate but higher computation cost at each iteration. Each node repeats this process until reaching a consensus across the network. 

\begin{wrapfigure}{r}{0.49\textwidth}
    \centering
    \includegraphics[width=0.99\linewidth]{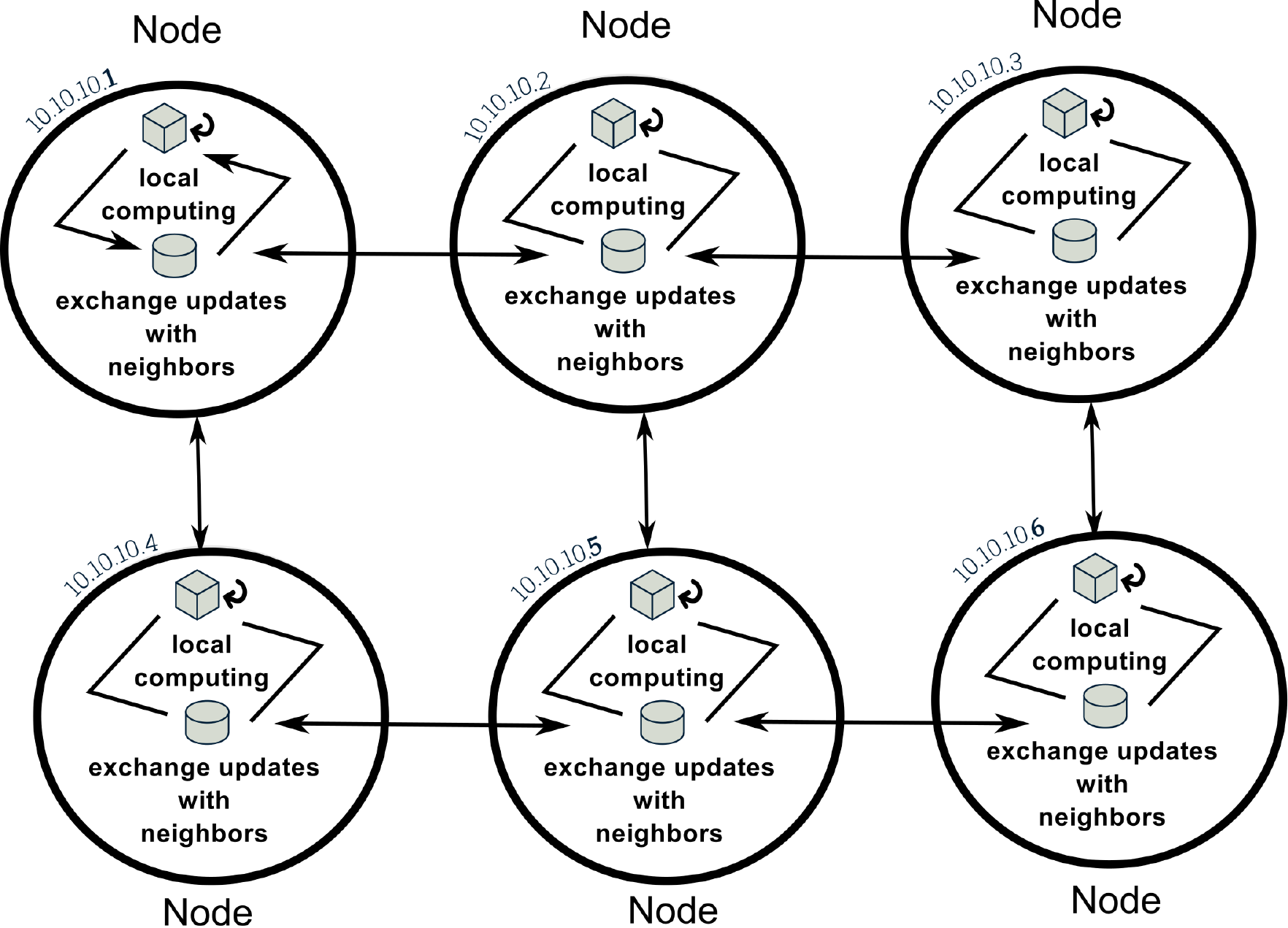}
    \caption{\textbf{Distributed iterative computing paradigm.}
    }
    \label{fig:decentralized}
\end{wrapfigure}
Distributed iterative computing is a paradigm-shifting computing problem and has received much attention \cite{SHY-UCLA2017,PXYY-UCLA2016, PXYY-SSC2016,AYSS-TSP2009} in computer science, mathematics \& statistics and machine learning community in the past five years. It is advancing rapidly, because it is increasingly necessary for many big data and Internet of Things applications, beyond subsurface imaging. In this new computing paradigm, each node holds an objective function privately known and can only communicate with its immediate neighbors (avoid multi-hop if possible). A great effort has been devoted to solving decentralized (fully distributed) consensus optimization problem, especially in applications like distributed machine learning, multi-agent optimization, etc. Several algorithms have been proposed for solving general convex and (sub)differentiable functions. By setting the objective function as least-square, the decentralized least-square problem can  be seen as a special case of the following problem:
\begin{equation}\label{equ:distributedopt}
\underset{x \in \mathbb{R}^n}{\min}  ~ F(x) := \sum_{i=1}^{p} F_i(x)
\end{equation}
where $p$ nodes are in the network and they need to collaboratively estimate the model parameters $x$. Each node $i$ locally holds the function $F_i$ and can communicate only with its immediate neighbors. Figure \ref{fig:decentralized} illustrates this paradigm. The literatures can be categorized into two categories:
(1) \textbf{Synchronous algorithms}, where nodes need to synchronize the iterations. In other words, each node needs to wait all its neighbors' information in order to perform the next computation round. Considering the problem in \eqref{equ:distributedopt}, (sub)gradient-based methods have been proposed \cite{Matei:2011a,Nedic:2009b,Nedic:2013a,Nedic:2014a,Chen:2012c,Jakovetic14}. However, it has been analyzed that the aforementioned methods can only converge to a neighborhood of an optimal solution in the case of fixed step size \cite{Yuan:2013a}. Modified algorithms have been developed in \cite{Chen:2012c} and \cite{Jakovetic14}, which use diminishing step sizes guarantee convergence. Other related algorithms were discussed in \cite{Zargham:2012a,Xiao:2005a,Tsitsiklis:1986a,Tsitsiklis:1984a,Terelius:2011a,Shi:2012a,Rabbat:2004a}, which share similar ideas.   
The D-NC algorithm proposed in ~\cite{Jakovetic14} was demonstrated to have an an outer-loop convergence rate of $O(1/k^2)$ in terms of objective value error. The rate is same as the optimal centralized Nesterov's accelerated gradient method and decentralized algorithms usually have slower convergence rate than the centralized versions. However, the number of consensus iterations within outer-loop is growing significantly along the iteration. Shi \cite{Shi14} developed a method based on correction on mixing matrix for Decentralized Gradient Descent (DGD) method \cite{Yuan:2013a} without diminishing step sizes.
(2) \textbf{Asynchronous algorithms}, where nodes do not need to synchronize the iterations. Decentralized optimization methods for asynchronous models have been designed in \cite{Wei:2013a,Iutzeler:2013b,Tsitsiklis:1986a}. The works in \cite{Wei:2013a,Iutzeler:2013b} leverage the alternating direction method of multipliers (ADMM) for the computation part, and in each iteration, one node needs to randomly wake up one of its neighbors to exchange information. However, the communication schemes in these two works are based on unicast, which is less preferable than broadcast in wireless sensor networks. Tsitsiklis \cite{Tsitsiklis:1986a} proposed an asynchronous model for distributed optimization, while in its model each node maintains a partial vector of the global variable. It is different from our goal of decentralized consensus such that each node contains an estimate of the global common interest. 
The first broadcast-based asynchronous distributed consensus method was proposed in \cite{AYSS-TSP2009}. However, the algorithm is designed only for consensus average problem without ``real objective function''. Nedic \cite{asyn_broadcast11} first filled this gap by considering general decentralized convex optimization similar as \eqref{equ:distributedopt} under the asynchronous broadcast setting. It adopted the asynchronous broadcast model in \cite{AYSS-TSP2009} and developed a (sub)gradient-based update rule for its computation. By replacing (sub)gradient computation with full local optimization, an improved algorithm has been designed in terms of the number of communication rounds \cite{zhao2017asynchronous}. 

For either synchronous or asynchronous algorithms, the design goal shall be to generate same or near-same results as centralized algorithm with minimal communication cost. The research on distributed iterative computing advances rapidly in the past several years. This section does not intend to survey all methods, but to merely point out some related works and potential direction on computing layer design for SAMERA. 

\subsection{Control Layer}
A control software with GUI can be connected to this network to view the computed 2D/3D images and adjust system and algorithm parameter settings such as resolution, filter and regularization parameters. The sensing, processing and computing layers can execute automatically and autonomously; on the other hand, the control layer allows users to control those layers, such as choose different parameters even different algorithm combinations, to achieve the desired effects. For example, user may choose to use migration imaging vs travel-time tomography based on the sensitivity to different types of seismic waves from subsurface properties and the SNR, or the combination with ambient noise imaging to view more or less details at the tradeoff of resource usages. This layer is currently application specific and depends on user preferences, but the user interface standard will gradually emerge in the future. 

\section{System Prototype Design Examples}
\label{sec:prototype}

This section will introduce several SAMERA system prototype design examples based on popular seismic imaging methods. In each of the following sections, the presentations of processing and computing layers will be emphasized, as they are the main intellectual challenges. The sensing and control layer are more or less an engineering and interface issues as described in the previous section.


\subsection{Travel-time Seismic Tomography}
\label{sec:tomott}
Travel-time seismic tomography (TomoTT) uses body wave (P and S) arrival times at sensor nodes to derive the subsurface velocity structure; the tomography model is continuously refined and evolving, as more seismic events are recorded over time. This method is often used in earthquake seismology, where the event source is a natural or injection induced (micro-)earthquake. TomoTT applies to the scenario where the signal-to-noise ratio (SNR) of body waves is good enough for arrival time picking.
Body wave arrival time picking and tomographic inversion are performed in processing and computing layers respectively. 

\begin{figure}[htb!]
\centering
    \includegraphics[width=0.99\textwidth]{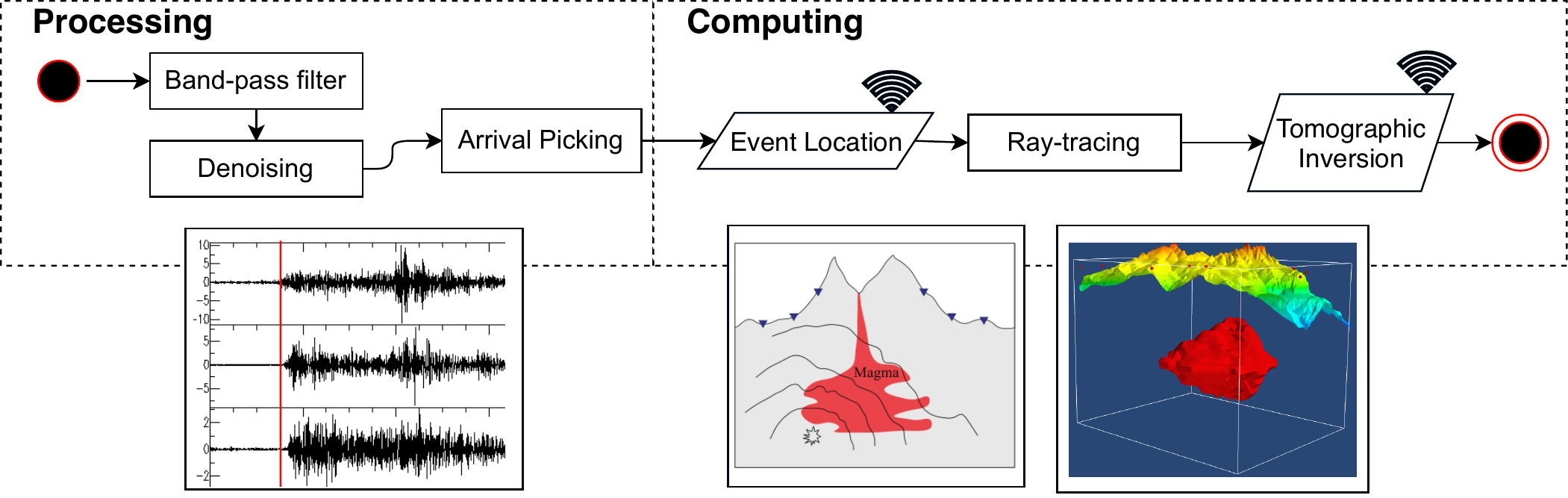}
    \caption{\textbf{Processing and computing algorithm flows of travel-time seismic tomography. The wireless sign in the figure means communication between nodes is needed.} 
    }
    \label{fig:tomott_flow}
\end{figure}


\subsubsection{Processing Layer}

When the SNR is relatively acceptable, the arrival time picking techniques are used to identify the body waves (P or S) onset times. Because of the strong background noise in seismic data~\cite{wong2009automatic,du2000noise}, the arrivals are hard to pick or even unidentifiable. The commonly used arrival picking algorithms are commonly based on the statistical anomaly detection methods, including, but not limited to, characteristic function (CF)~\cite{allen1978automatic}, the short- and long-time average ratio (STA/LTA)~\cite{baer1987automatic}, Akaike information criterion (AIC)~\cite{takanami1991estimation}, wavelet transform (WT)~\cite{anant1997wavelet}, cross-correlation~\cite{molyneux1999first}, modified energy ratio (MER)~\cite{wong2009automatic}, higher order statistics (HOS)~\cite{yung1997example,li2014automatic}, and so on. 
In general, the arrival picking~\cite{akram2016review} problem can be formulated as the ratio of short-term characteristic function and long-term characteristic function (STCF/LTCF).
The characteristic function (CF)~\cite{allen1978automatic} can be some kind of statistical metrics, including energy~\cite{wong2009automatic}, moments~\cite{li2014automatic} and likelihood estimates~\cite{li2016online}.
	
Most arrival picking methods were designed based on single channel (e.g., vertical axis) seismic data. With triaxis geophones or three components seismic data, two polarization parameters can be used to distinguish between P and S waves, and noise~\cite{baillard2014automatic}, because the wave types and orientations affect the polarization of signal onsets. As the data have three orthogonal ground-motion records corresponding to E, N, and Z, the onset polarization could indicate the types (surface or body), phase (P or S) of the wave~\cite{li2017automatic}.	


\subsubsection{Computing Layer}

The picked arrival times are then used to estimate the event source
location and origin time in the subsurface, as shown in Fig.~\ref{fig:tomott_flow}. Thereafter, the ray-tracing and tomography inversion will be performed. Given the source locations of the seismic events and initial velocity model, ray tracing is to find the ray paths from the seismic source locations to the sensor nodes. 
After ray tracing, the seismic tomography problem is formulated as a large sparse matrix inversion problem.
Suppose there are total $M$ seismic events and $N$ sensors, and  $L$ cells in the 3D tomography model, 
then let $\mat{A} \in \mathbb{R}^{N\dot M\times L}$ be the matrix of ray information between $M$ events and $N$ sensors, $\vec{t} \in \mathbb{R}^{N\dot M \times 1}$ be the vector of travel time between $M$ events and $N$ sensors, and $\vec{s} \in \mathbb{R}^{L \times 1}$ be the 3D tomography model to calculate. The tomographic inversion problem can be formulated as  
\begin{equation}
\vec{s}^* = \arg\min_{\vec{s}} \| \vec{t} - \mat{A} \vec{s} \|_2^2 + \lambda \| \vec{s} \|_2^2
\label{eq:ls}
\end{equation}
where $\lambda$ be the regularization parameter.
In centralized algorithm, the system of equations is solved by sparse matrix methods like LSQR 
or other conjugate gradient methods \cite{lees1991bayesian}.
Various parallel algorithms have also been developed to speed up the execution of these methods \cite{urdaneta82shortest,liu2006parallel,butrylo2006distributed}. However, designed for high-performance computers, these centralized approaches need significant amount of computational/memory resources and require the global information (e.g., $\vec{t}$ and $\mat{A}$). 

By the way, double-difference tomography \cite{zhang2006development} was developed to simultaneously solving the event location and three-dimensional tomography model. It claims to produce more accurate event locations and velocity structure near the source region than standard tomography. Its mathematical formulation is in the same format as equation \ref{eq:ls}. 

\subsubsection{Compute TomoTT in Sensor Networks}
To implement travel-time seismic tomography in sensor networks, an effective approach is to let each node compute tomography in an asynchronous fashion and communicate with neighbors only. By eliminating synchronization point and multi-hop communication that are used in existing distributed algorithms, the system can scale better to a large number of nodes, and exhibit better fault-tolerance and stability. In the harsh geological field environment, the network disruptions are not unusual and reliable multi-hop communication is not easy to achieve.
Prototype system based on TomoTT \cite{song2015real,kamath2016distributed,zhao2015decentralised,kamath2017distributed,ramanan2015indigo,lei2016sensor,kamath2015dristi,kamath2015distributed,kamath2013component,shi2013imaging} was designed and demonstrated. 
In~\cite{shi2013imaging}, a distributed computing algorithm based on vertical partition was proposed. The key idea is to split the least-square problem into vertical partitions, similar as multisplitting method but being aligned with the geometry of tomography. Later a nodes in each partition is chosen as a landlord to gather necessary information from other nodes in the partition and compute a part of tomography. The computation on each landlord is entirely local and the communication cost is bounded. After the partial solution is obtained, it is then combined with other local solutions to generate the entire tomography model. In ~\cite{kamath2013component,kamath2015dristi,kamath2016distributed}, the block iterative kaczmarz method with component averaging mechanism~\cite{GG-SIAM2005,CGG-PC2001} was proposed. The key idea is that each nodes runs multiple iterations of randomized kaczmarz, then their results are aggregated through component averaging then distributed back to each node for next iterations. After multiple iterations, the algorithm converges and generates the tomography. Decentralized synchronous and  synchronous methods with random gossip and broadcast \cite{zhao2015decentralised,zhao2017asynchronous,zhao2015fast} were also developed to solve the inversion problem in equation \ref{eq:ls}.

\subsection{Migration-based Microseismic Imaging}
\label{sec:migration}
\begin{figure}[htb!]
\centering
    \includegraphics[width=0.99\textwidth]{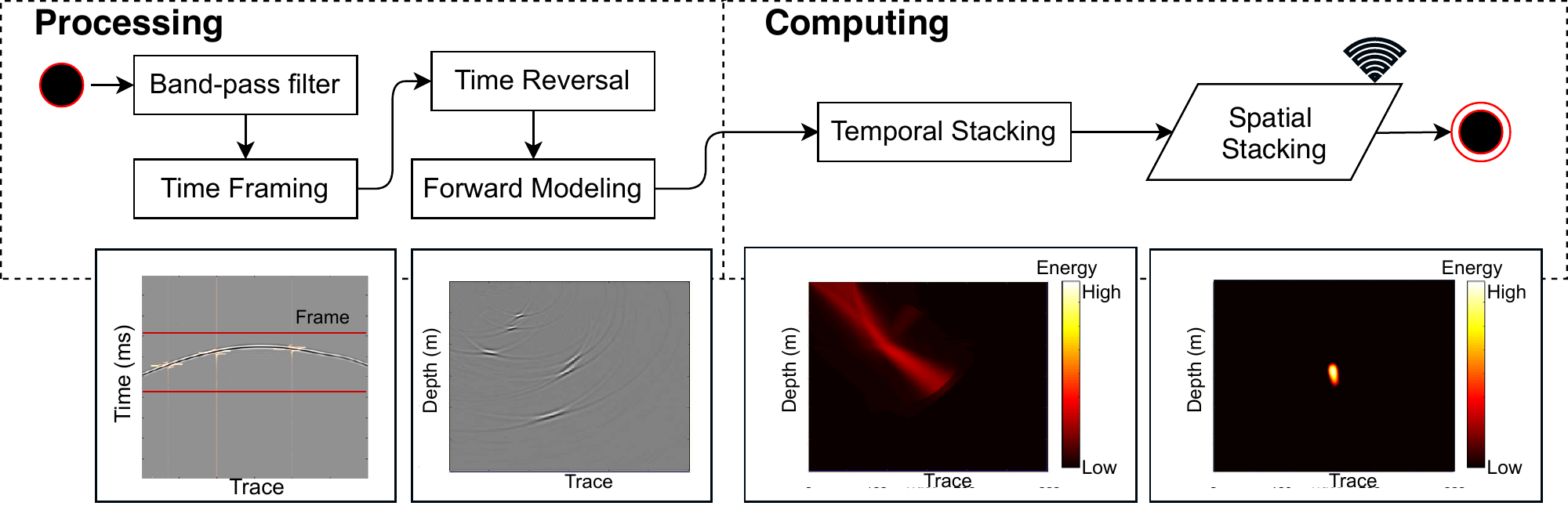}
    \caption{\textbf{Processing and computing algorithm flows of seismic migration imaging. The wireless sign in the figure means communication between nodes is needed.}
    }
    \label{fig:mmi_flow}
\end{figure}

Migration-based Microseismic Imaging (MMI) applies the reverse-time migration (RTM) principles to locate the microseismic source locations~\cite{artman2006imaging,wong2015imaging,sun2015investigating,nakata2016reverse}. With a given velocity model, the time-reversed extrapolation of the observed wavefields can be calculated based on wave equations. The extrapolated wavefields from different receivers stack together to enlighten the location of seismic sources~\cite{wu2017microseismic}, as illustrated in Figure \ref{fig:mmi_flow}. MMI~\cite{kaoshan04,gajewski05,mcmechan82,atm10,witten11,kremers11,wu2017microseismic} typically uses body waves and has two main steps: forward modeling and stacking (also called imaging condition) in processing and computing layer, respectively. 



\subsubsection{Processing Layer}

In the first step, a bandpass filter with an appropriate bandwidth shall be applied, which is narrow enough to contain signals of interest, and not too narrow to filter out signal components. For active source migration, since the sources are controllable and seismic acquisition array density is high, the raw data and the corresponding wavefields can be completely separated for specific source characterization\cite{yavuz2008space,wong2015imaging}. While, for passive seismic source imaging, the determination of a seismic event becomes critical, or else, there will be not enough time series data for used for wavefield construction or computation resources are wasted on background noises~\cite{sun2015investigating,nakata2016reverse,yang2016time}. To extract the window of seismic signal segments containing events, time framing based on energy segmentation can be applied (Fig.~\ref{fig:mmi_flow}). 
In addition, for better location, a localized normalization operator is necessary to deal with the issue of unbalanced amplitudes~\cite{sun2015investigating}.


Let $S(\mathbf{x};t;\mathbf{x_s})$ denote the source wavefield generated from the source location $\mathbf{x_s}$ and recorded at a spatial location $\mathbf{x}$ following the wave equation $\left( \frac{1}{v^2(\mathbf{x})} \frac{\partial^2}{\partial t^2} - \nabla^2\right) S(\mathbf{x};t;\mathbf{x_s}) =0$, where $v$ is velocity, and $\nabla^2$ is the (spatial) Laplacian operator~\cite{zhang2007true}.  RTM algorithms use the zero lag of the crosscorrelation between the source and receiver wavefields to produce an image $I_{r_{i}}$ at the receiver location $\mathbf{x}_{r_i}$~\cite{claerbout1985imaging,wong2015imaging}:
\begin{equation}
I_{r_{i}}(\mathbf{x},t) = S(\mathbf{x};t;\mathbf{x_s}) R_i(\mathbf{x};t;\mathbf{x_s})
\label{eq:rtm}
\end{equation}
Here $R_i(\mathbf{x};t;\mathbf{x_s})$ is the receiver wavefield, which is approximated using a finite-difference solution of the wave equation~\cite{zhang2007true,araya2011assessing,liu2017two}. For microseismic imaging, under the virtual source assumption, the source wavefield is eliminated by seismic interferometry using receiver wavefields~\cite{schuster04,nakata2016reverse,wu2017microseismic}.

\subsubsection{Computing Layer}
{Eq. (\ref{eq:rtm}) infers that every receiver  generates a 4D wavefield $I_{r_{i}}$. The imaging condition step is to combine all receivers' wavefields to form the final migration images. It is often produced by summation of wavefields: 
\begin{equation}
\label{eq:ic-tradition}
I(\mathbf{x},t) = \sum\limits_{i=0}^{N-1} I_{r_{i}}(\mathbf{x},t)
\end{equation}
Conventionally, this is done by backward-propagating all the data from all sensors at once. Assuming the velocity model is accurate and data contains zero noise, the image $I(\mathbf{x},t)$ should have non-zero values only if all the backward-propagated wavefields are non-zero at the seismicity location $\mathbf{x}$ and time $t$. However, it does not work well with real data with noises. Hybrid imaging condition \cite{sun2015investigating} was proposed for more effective microseismic imaging as described in equation \ref{eq:ic-tradition}:
\begin{equation}
\label{eq:ic-hyb}
I(\mathbf{x},t) = \prod\limits_{j=0}^{N/(n-1)} \sum\limits_{k=0}^{n-1} I_{r_{j \times n+k}}(\mathbf{x},t)
\end{equation}
where $n$ is the local summation window length. Length $n$ should be selected such that neighboring receivers are backward-propagated together while far-apart receivers are cross-correlated. Equation~\ref{eq:ic-hyb} requires $N/n$ computations of reverse-time modeling. Notice that the hybrid imaging condition parameter decision needs a careful design and evaluation by considering the trade-off between network resource constraints and image quality. 

This method is capable of producing high-resolution images of multiple source locations. It adopts the migration imaging principles for locating microseismic hypocenters. It treats the wavefield back-propagated from each individual receiver as an independent wavefield, and defines microseismic hypocenters as the locations where all the wavefields coincide with maximum local energies in the final image in both space and time. 
Microseismic monitoring based on migration imaging is currently considered as the most effective technique to track the geometry of stimulated fracture networks in resource extraction \cite{maxwellbook}. 

\subsubsection{Compute MMI in Sensor Networks}
In sensor networks, the temporal stacking and spatial stacking in  Fig.~\ref{fig:mmi_flow} can be implemented based on equation~\ref{eq:ic-hyb-collapse}, which is a slight modification from equation \ref{eq:ic-hyb}.  
\begin{equation}
\label{eq:ic-hyb-collapse}
I(\mathbf{x}) = \prod\limits_{j=0}^{N/(n-1)} I_{c_{j}}(\mathbf{x}) = \prod\limits_{j=0}^{N/(n-1)} \sum_t\sum\limits_{k=0}^{n-1} I_{r_{j \times n+k}}(\mathbf{x},t)
\end{equation}
In equation~\ref{eq:ic-hyb-collapse}, $I_{r_{j \times n+k}}(\mathbf{x},t) $ is the 4D wavefield from each node, and $I_{c_{j}}(\mathbf{x})$ is the 3D temporal stacking image of each cluster. In other words, it is a dimension reduction operation from 4D to 3D. The temporal stacking including summation of wavefields and time axis collapses can be performed in a cluster of sensors~\cite{witten2015extended}. This decreases the communication cost in the next step, where the spatial stacking is performed between clusters. In spatial stacking, the images of the same location $\mathbf{x}$ from different clusters are essentially cross-correlated. The communication cost of passing the 3D image $I_{c_j}(\mathbf{x})$ is still considered expensive for sensor networks. Gaussian beam migration can be further applied to limit the computation and communication to a narrow beam ~\cite{rentsch2004location}, instead of full wavefield $I_{r_{j \times n+k}}(\mathbf{x},t)$. A primitive prototype of SAMERA on migration-based microseismic imaging has been designed \cite{wang2018microseismic}. 
By using Gaussian beams around these rays, the stacking of amplitudes is restricted to physically relevant regions only. This reduces tens of times of computational and communicational burden, without damaging the imaging quality. 

\subsection{Ambient Noise Seismic Imaging}
\label{sec:ansi}

\begin{figure}[htb!]
\centering
    \includegraphics[width=0.99\textwidth]{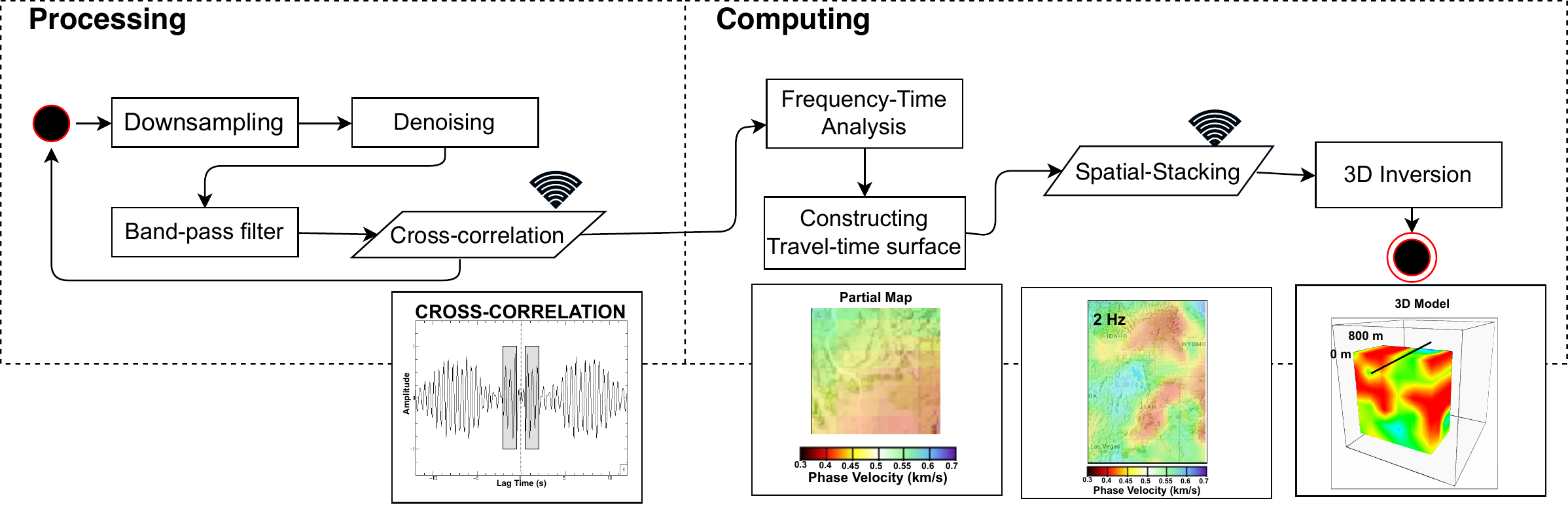}
    \caption{\textbf{Processing and computing algorithm flows of ambient noise seismic imaging. The wireless sign in the figure means communication between nodes is needed.}
    }
    \label{fig:ansi_flow}
\end{figure}

To fully utilize the dense seismic array when there are few earthquakes or active sources, Ambient Noise Seismic Imaging (ANSI)~\cite{Lin2008,Moschetti2010,Lin2011} has been developed to image subsurface using surface waves. ANSI uses radiation from random sources in the earth to first estimate the Greens function between pairs of stations \cite{roux2004,Snieder2004,Shapiro2005,Gouedardetal2008} and then invert for a 3D earth structure ~\cite{Moschetti2010,Lin2011}. 
Many applications have relied on relatively low frequency data (between 0.05-0.5 Hz) from the ocean noise \cite{longuet1950theory}, which images structure in the scale of kilometers, while more local structures can be imaged with higher order wave modes (higher frequencies) and denser networks \cite{Picozzi2009,Yang2011,Lin2013}.  
The main algorithm flows of ANSI (Figure~\ref{fig:ansi_flow}) are described as follows:

\subsubsection{Processing Layer}
This layer is to derive surface wave travel times through noise cross-correlations. Data conditioning will be first applied to the raw seismic signals, such as downsampling, denoising, band-pass filtering. Thereafter, the noise cross-correlation $C_{AB}$ between two stations will be performed:
\begin{equation}
C_{AB}(t) = \int_{-\infty}^\infty u_A(\tau) u_B(t+\tau) d\tau = \int_{-\infty}^\infty [-G_{AB}(\tau) + G_{AB}(-\tau) ]d\tau .
\label{eq:correlation}
\end{equation}
where $u_A$ and $u_B$ are the recorded noises at locations A and B ~\cite{Bensen2007}. Theoretical studies have shown that if the noise wavefield is sufficiently diffusive, the cross-correlation between two stations can be used to approximate the Green's function $G_{AB}$ between the two sensors or locations ~\cite{Lobkis2001,Snieder2004}. By calculating ambient noise cross-correlations between one center station and all other stations, seismic wavefield excited by a virtual source located at the center station can be constructed. Based on the noise cross-correlations, the period dependent surface wave phase and group travel time can be determined between each pair of stations. 

\subsubsection{Computing Layer}
This layer first generates a series of frequency dependent 2D surface wave phase velocity maps and then 3D inversions are performed across the array to form the final 3D tomography. The eikonal and Helmholtz tomography methods will be adopted to determine 2D phase velocity maps based on empirical wavefield tracking ~\cite{Lin2009,Lin_Ritz2011}. For each event $i$, it measures surface wave phase velocities at each location directly by the spatial derivatives of the observed wavefield:
\begin{equation}
\frac{1}{c_i^2({\bf r})}  =|\nabla \tau({\bf r}_i, {\bf r})|^2 - \frac{\nabla^2 A_i({\bf r})}{A_i({\bf r}) \omega^2},
\label{wavefield}
\end{equation}
where $\tau$ and $A$ represent phase travel time and amplitude measurements, and $\widehat{\bf k}_i \cong \nabla \tau ({\bf r}_i, {\bf r}){c_i({\bf r})}$, $c$ and $\omega$ are direction of wave propagation, phase velocity and angular frequency, respectively.  $\widehat{\bf k}_i$ can be derived directly by solving 2D Helmholtz wave equation also called eikonal equation, can be derived from equation~\ref{wavefield} under infinite frequency approximation. While the above equations are defined for `events,' it's important to note that the cross-correlation method from equation~\ref{eq:correlation} effectively turns each station into an `event' recorded at every other station, and so the wavefield from virtual sources at each station, as well as the spatial derivatives in equations~\ref{wavefield} can be approximated from the set of cross-correlations with that station.
After the 2D image reconstruction, the frequency dependent phase velocities at each location can then be used to invert for vertical profiles. The combination of all 2D model and vertical profile across the study area produces the final 3D model. 

\subsubsection{Compute ANSI in Sensor Networks}
To compute ANSI in sensor networks, there are two main challenges and can be addressed as follows:
(1) The noise cross-correlation step will require every pair of nodes to exchange data with each other at the beginning. The communication cost can be reduced by subsampling, applying a bandpass filter, and limiting the cross-correlation between nodes in near-range while approaching the far-range cross-correlation through distributed interpolation.
(2) The eikonal tomography step will require all nodes to stack their locally calculated velocity maps to form the final 2D/3D subsurface image. The stacking processing can be done through in-network data aggregation or decentralized consensus. Each approach has its own advantages and disadvantages: aggregation works better when the network is reliable, while consensus might be better when network is intermittent. A primitive prototype of SAMERA based on ANSI \cite{valero2017real} has been built and demonstrated.


\section{Conclusion}
Creating SAMERA requires an interdisciplinary collaboration and advancement of sensor networks, signal processing, distributed computing, and geophysical imaging. Prototypes based on several seismic imaging methods have been demonstrated, yet there are many research challenges and opportunities, such that: (1) Fully automation: the geophysical imaging today often involves human in the loop, which is not a surprise as the first generation optical camera is not fully automatic too. For example, the initial velocity model and some algorithm parameters are still based on experience, so the questions are: how to self-learn and self-optimize parameters to make subsurface imaging computing fully automatic? how to build initial velocity model automatically? how to integrate machine learning (e.g., data-driven) with physics-based modeling to enable better automation? (2) Fast completion: distributed iterative computing frameworks and principles have been laid out, yet many practical issues to address, such as: how to decide stopping/pausing criteria to avoid over-fitting? how to generate subsurface image faster under the bandwidth constraint and random network failures? (3) Data fusion: different geophysical sensors/methods are sensitive to different geophysical properties of subsurface, how to integrate different geophysical methods for joint inversion to generate better subsurface images? Those research questions are no longer unique to SAMERA creation and have received much attention in big data, machine learning, Internet of Things and other domains, beyond geophysics, better and better solutions are developed everyday. With those recent rapid advancement and cost reduction of both hardware and computing algorithms, it is the right time to start creating SAMERA - the camera to see through the subsurface is coming! 

\newpage

\bibliography{sensorweb}

\newpage
\section*{Authors Contribution Statement}
The following statement of responsibility specifies the contribution of every author of the paper ``Toward Creating Subsurface cAMERA: SAMERA''. The corresponding author was the main contributor of the paper. The other three authors had contributed equally to the work. All authors reviewed the manuscript.

\subsection*{WenZhan Song (Corresponding author)}
The author envisioned the idea of SAMERA and the main concepts and fundamental theory behind the idea. He wrote the main structure of the manuscript text, reviewed the manuscript, and was responsible of the well-explanation of the manuscript structure. 

\subsection*{Fangyu Li}
The author contributed to the writing of sections 3.1 and 3.2. 

\subsection*{Maria Valero}
The author contributed to the writing of section 3.3. Author also contributed with the preparation of all the figures in the manuscript.

\subsection*{Liang Zhao}
The author contributed to the writing of section 2.3.

\section*{Additional Information}

\textbf{Competing Interests Statement}

The authors (WenZhan Song, Fangyu Li, Maria Valero, Liang Zhao) certify that they have NO affiliations with or involvement in any organization or entity with any financial interest (such as honoraria; educational grants; participation in speakers’ bureaus; membership, employment, consultancies, stock ownership, or other equity interest; and expert testimony or patent-licensing arrangements), or non-financial interest (such as personal or professional relationships, affiliations, knowledge or beliefs) in the subject matter or materials discussed in this manuscript.

\end{document}